# Two new mathematical models to predict the flow stress at hot deformation[*]

Soheil Solhjoo[†]

Department of Mathematics and Natural Sciences, University of Groningen, Groningen, The Netherlands

Abstract

Based on both linear and non-linear estimations of work hardening rate versus strain curves, two mathematical models have been developed to predict the flow curves under hot working conditions up to the peak. The models were tested for a mechanically alloyed Al6063/0.75Al2O3/0.75Y2O3 nanocomposite under different hot forming conditions. The predicted results from both models are found to be in accord with the experimental flow stress curves. However, the linear model (with an average error of 0.81%) predicted the flow stress more accurate than the non-linear model (with an average error of 1.94%).

## 1. Introduction

Most of the models of the flow stress curves divide the stress-strain curves into two regions; the first one involves the effects of work hardening and dynamic recovery on flow stress and the second one considers the

---

softening caused by dynamic recrystallization. Some of these models treat the whole curve with only one equation, e.g. the well-known sine hyperbolic equation proposed by Garofalo [1].

In this investigation two new models are introduced which can be used for estimation of the flow stress curves of metals and alloys up to the peak under hot deformation processes. The analyses have been done to determine the flow stress values of mechanically alloyed Al6063/0.75Al2O3/0.75Y2O3 nanocomposite. The calculated flow stresses are in accord with the experimental ones.

## 2. Modeling of flow stress curves up to peak

The slope of the flow stress curve determined at a constant strain rate and temperature corresponds to the work hardening rate, i.e. $\theta = \dfrac{d\sigma}{d\varepsilon}\bigg|_{T,\dot\varepsilon}$, where $\dot\varepsilon$ and $T$ are strain rate and absolute temperature, respectively.

In this investigation both the linear and non-linear estimations of the work hardening rate vs. strain are taken into account in order to build up two formulations for the flow stress up to the peak.

### 2.1. Linear estimation

It is known that the linear estimation of $\theta - \varepsilon$ curve could be used to develop a constitutive equation to estimate the flow stress up to the peak [2]. This linear estimation would be in the form of:

$$\frac{d\sigma}{d\varepsilon} = A\varepsilon + C \tag{1}$$

where $A$ is the slope of the line, and $C$ is a constant. Using the root of the $\theta - \varepsilon$ curve ($\varepsilon = \varepsilon_P, \theta = 0$), the value of $C$ is obtained to be $-A\varepsilon_P$. Previously, I solved this differential equation [2]. However, the constant of the integral was omitted to give the formula the ability of predicting the critical strain for the initiation of dynamic recrystallization. This omission made the model unable to accept an initial value ($\sigma_0$) unless with some additional modifications [3]. It should be noted that Eq.1 can be solved by keeping the constant of integral, which results in:

$$\sigma = \sigma_P - \frac{B}{2}(\varepsilon_P - \varepsilon)^D \qquad (2)$$

where $\sigma_P$ and $\varepsilon_P$ are stress and strain corresponding to the peak, respectively. Analytically, the exponent $D$ is determined to be 2 (see "Appendix A"); however, considering it as a variable would be more acceptable. Using the initial point ($\varepsilon = 0, \sigma = \sigma_0$) the constant $B$ is found to be $2\frac{\sigma_P - \sigma_0}{\varepsilon_P^D}$. It means that with no further modification, the initial stress is a part of the model. By replacing the value of $B$ in Eq.2, it can be rewritten as follows:

$$\sigma = \sigma_P - (\sigma_P - \sigma_0)\left(1 - \frac{\varepsilon}{\varepsilon_P}\right)^D \qquad (3)$$

2.2. Non-linear estimation

A non-linear estimation of $\theta - \varepsilon$ curve up to the peak could be expressed by the following equation:

$$\frac{d\sigma}{d\varepsilon} = \frac{G}{\varepsilon} - F\varepsilon \qquad (4)$$

where $G$ and $F$ are constants. Using the root of the $\theta - \varepsilon$ curve ($\varepsilon = \varepsilon_P, \theta = 0$), the value of $F$ is equal to $\frac{G}{\varepsilon_P^2}$

. Solution of the differential Eq.4 would be:

$$\sigma = \sigma_P + G\left[\ln\left(\frac{\varepsilon}{\varepsilon_P}\right) - \frac{1}{2}\left(\left(\frac{\varepsilon}{\varepsilon_P}\right)^2 - 1\right)\right] \qquad (5)$$

# 5. Numerical results

The experimental results of mechanically alloyed Al6063/0.75Al2O3/0.75Y2O3 nanocomposite from [4] are used for mathematical analysis of this investigation. The considered deformations are done from 300˚C up to 500˚C with steps of 50˚C, and three strain rates of 0.01, 0.1 and 1 s$^{-1}$.

## 5.1. Linear estimation

In order to estimate the flow stress using the linear estimation, the stress must be calculated from Eq.3. As Fig.1 shows, the linear plots of $\ln(\sigma_P - \sigma)$ vs. $\ln\left(1 - \dfrac{\varepsilon}{\varepsilon_P}\right)$ were used to determine the value of $D$ which is found to be 2.036. It should be noted that the x-intercept of this plot equals to $\ln(\sigma_P - \sigma_0)$ which is useful when $\sigma_0$ is unavailable. Therefore, this value could be used if the value of the initial stress is not accessible. Moreover, this problem could be solved using a known point with the smallest strain (see "Appendix B").

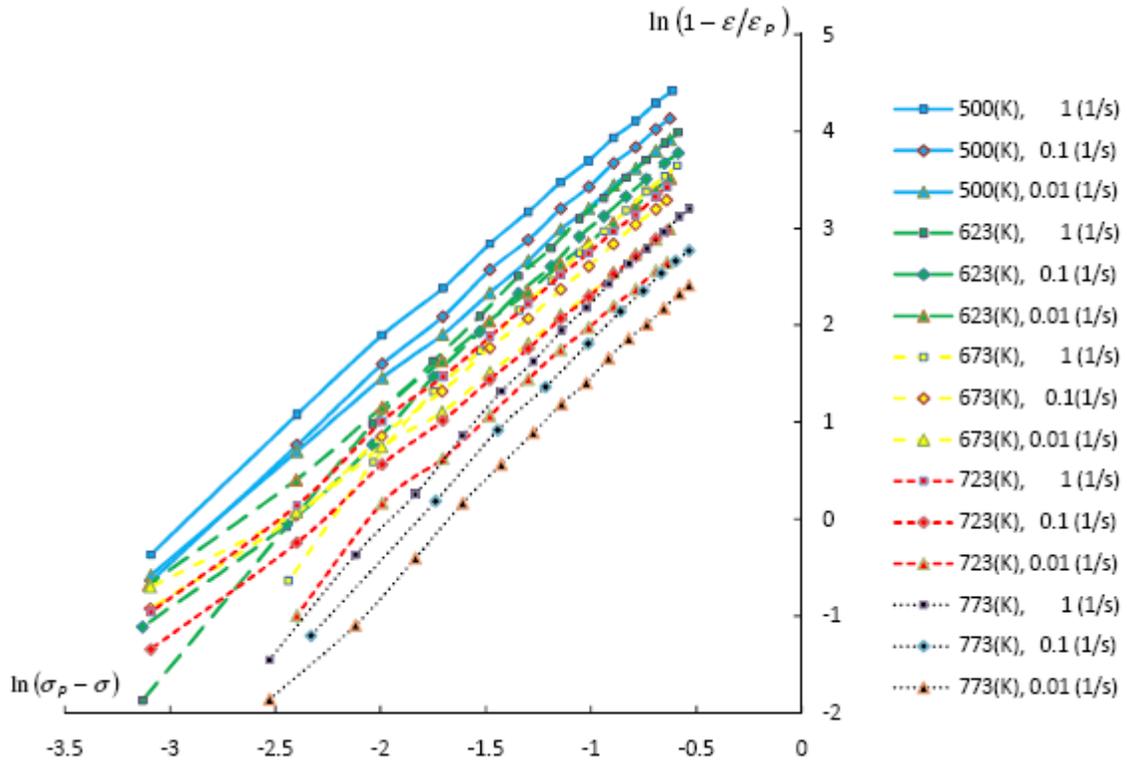

**Fig. 1.** The plots for evaluation of constant *D* of Eq. 3.

### 5.2. Non-linear estimation

The stress must be determined from Eq.5 in order to use non-linear estimation for the flow curves. Linear plots of $(\sigma - \sigma_P)$ vs. $\ln\left(\dfrac{\varepsilon}{\varepsilon_P}\right) - \dfrac{1}{2}\left(\left(\dfrac{\varepsilon}{\varepsilon_P}\right)^2 - 1\right)$ were used to determine the parameter *G*. This parameter is found to be a function of hot deformation conditions. The following formula [3] can be used to predict the values of *G*:

$$G = \overline{G}\dot{\varepsilon}^{E} \exp(\varphi T) \qquad (6)$$

where $\overline{G}$, *E* and $\varphi$ are constants. As it is explained in [3], the constants *E* and $\varphi$ should be calculated from $\ln \dot{\varepsilon}$ vs $\ln G$ (Fig.2) and $T$ vs $\ln G$ (Fig.3) plots, respectively. In this research, *G* could be calculated from Eq.7:

$$G = 17247.8\,\dot{\varepsilon}^{0.131}\exp(-0.0076\,T) \tag{7}$$

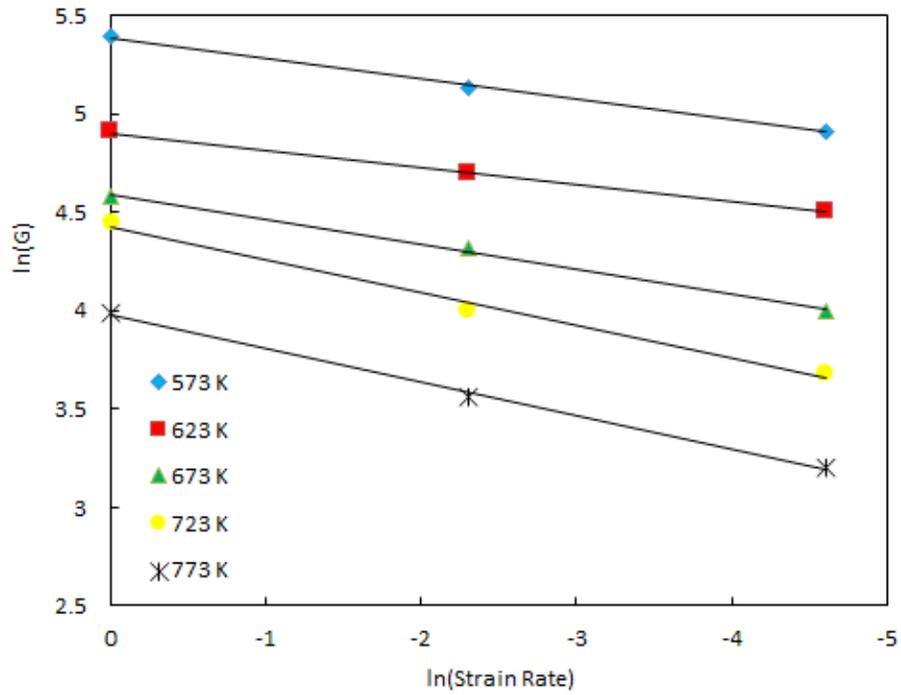

**Fig. 2.** Dependence of parameter *G* on strain rate.

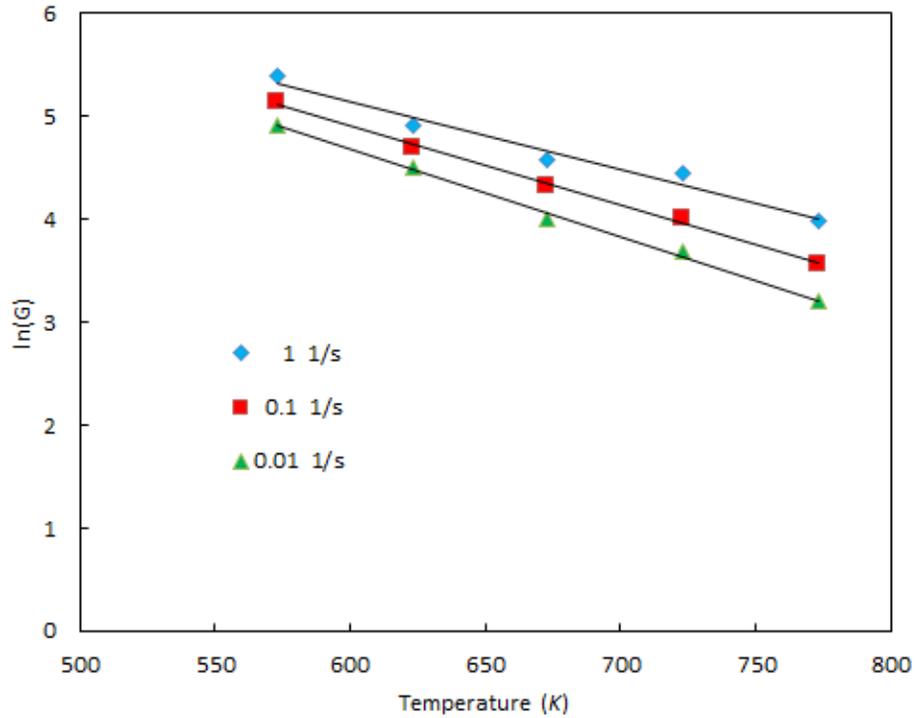

**Fig. 3.** Dependence of parameter *G* on temperature.

# 6. Discussion

The numerical solution of *D* was determined to be 2.036 which is highly in accordance with its analytical value (see "Appendix A"). In order to utilize either of these models, the values of stress and strain corresponding to the peak are needed. In this study, I used their available experimental measurements. However, the modeling of the peak stress is discussed in Appendix C. Besides, the value of $\sigma_P - \sigma_0$ is extracted from the x-intercept values of plots in Fig.1 for the estimation of the flow stress by Eq.3.

Figs.4 and 5 show the experimental flow stress vs. predicted ones by Eqs.3 and 5 (for more than 200 data point), respectively. The minimum, mean and maximum percent errors are listed in table 1.

Table 1. Percent errors of the predicted flow stress values

|      | Minimum | Average | Maximum |
|------|---------|---------|---------|
| Eq.3 | 0.00    | 0.81    | 2.86    |
| Eq.5 | 0.00    | 1.94    | 10.04   |

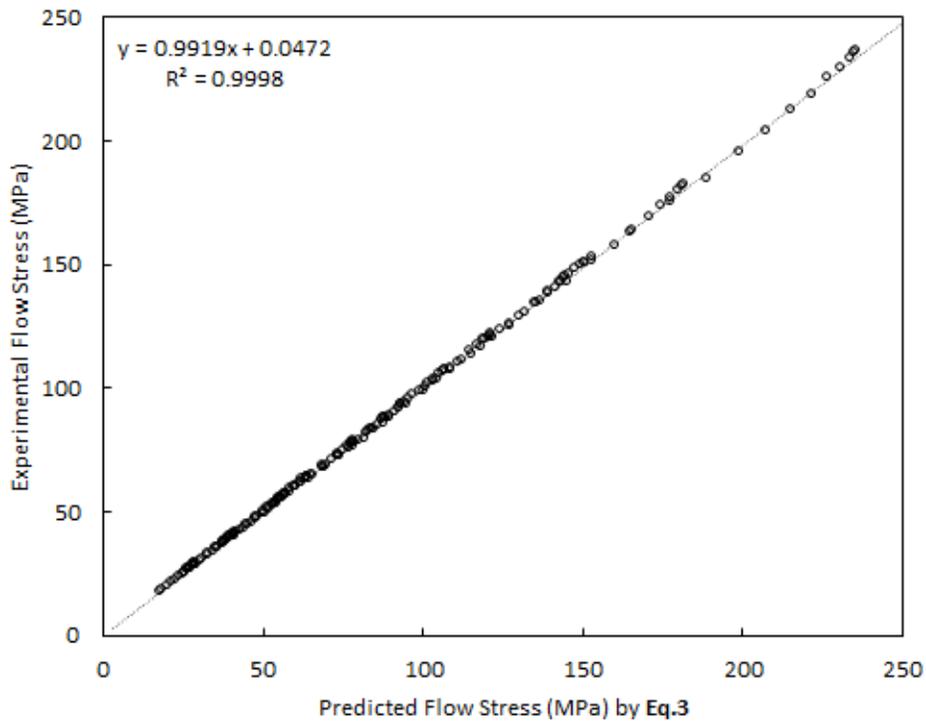

**Fig. 4.** Comparison between experimental and predicted flow stress by Eq. 3.

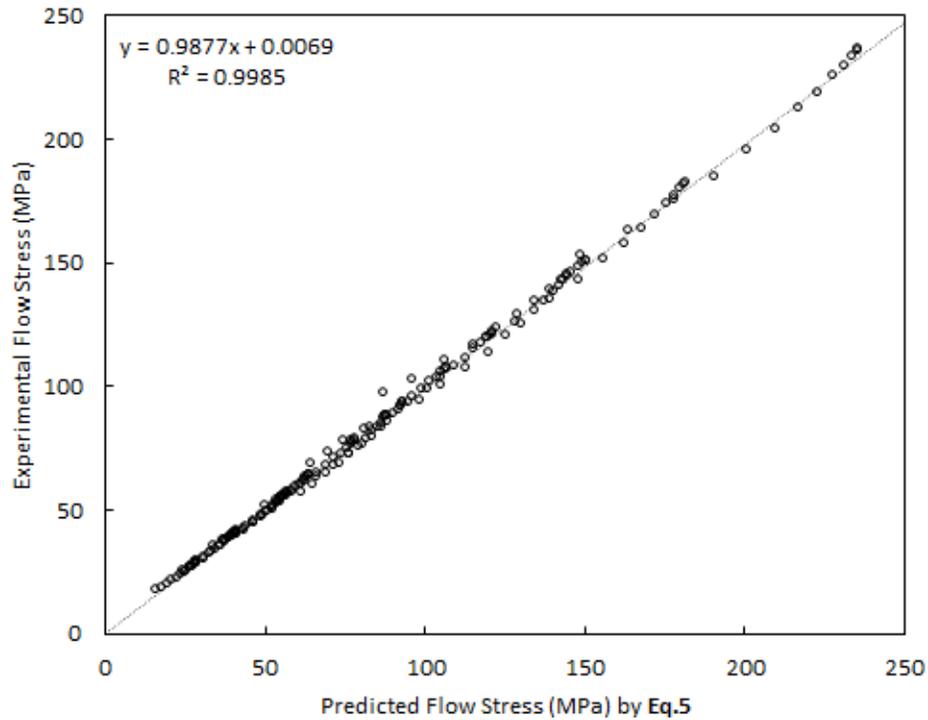

**Fig. 5.** Comparison between experimental and predicted flow stress by Eq. 5.

As it can be seen, both models show high correlation between empirical and predicted flow stress values. However, the error is smaller for Eq.3, which might indicate that a linear estimation of work hardening rate vs. strain curves would show better results than a non-linear estimation in the form of Eq.5.

It should be noted that utilizing the linear estimation (Eq.3) is also much easier than the non-linear one, because the constant of Eq.3 (*D*) is not a function of material or conditions of the deformation process. While, Eq.5 needs the values of *G* which should be precisely determined through heavy mathematical calculations.

# 7. Conclusion

Two new mathematical constitutive models of linear and non-linear estimation of work hardening rate as a function of strain are presented. The equations could model the flow stress curves up to the peak. Both models have been examined for mechanically alloyed Al6063/0.75Al2O3/0.75Y2O3 nanocomposite and results show the high accuracy of both models. It is also found that the linear estimation shows better results, and is easier to use.

# Appendix A

Starting with integration of Eq. 1, the solution would be:

$$\sigma = A\varepsilon\left(\frac{1}{2}\varepsilon - \varepsilon_P\right) + Const \qquad (1-A)$$

In order to determine the value of *Const*, we may use the value of stress and strain corresponding to the peak. Therefore, we have:

$$\sigma_P = A\varepsilon_P\left(\frac{1}{2}\varepsilon_P - \varepsilon_P\right) + Const \qquad (2-A)$$

Which results in:

$$Const = \sigma_P + \frac{1}{2}A\varepsilon_P^2 \qquad (3-A)$$

Using this value, Eq. (1-A) changes to:

$$\sigma = \sigma_P + \frac{1}{2}A\varepsilon_P^2 + \frac{1}{2}A\varepsilon^2 - A\varepsilon_P\varepsilon \qquad (4-A)$$

As the final step, using a simple polynomial identity in algebra, Eq. (4-A) could be rewritten as follows:

$$\sigma = \sigma_P + \frac{1}{2} A(\varepsilon_P - \varepsilon)^2 \qquad (5-A)$$

As Eq. (5-A) shows, it is a quadratic equation. However, as it is explained in the text, it would be safer to accept it as an adjustable parameter.

## Appendix B

It is possible that the initial stress is unknown. Therefore, there would be a problem in order to utilize the linear estimation (Eq.3). In such cases (like in the present study) we can use a known point and make the formula applicable. Let's call the stress and strain of the known point $\sigma_N$ and $\varepsilon_N$, respectively. Eq.3 would be written for this point as:

$$\sigma_N = \sigma_P - (\sigma_P - \sigma_0)\left(1 - \frac{\varepsilon_N}{\varepsilon_P}\right)^D \qquad (1-B)$$

The value of $\sigma_0$ is easily found to be:

$$\sigma_0 = \sigma_P - (\sigma_P - \sigma_N)\left(1 - \frac{\varepsilon_N}{\varepsilon_P}\right)^{-D} \qquad (2-B)$$

Considering Eq.(2-B), Eq.3 could be rewritten as follows:

$$\sigma = \sigma_P - (\sigma_P - \sigma_N)\left(\frac{\varepsilon_P - \varepsilon}{\varepsilon_P - \varepsilon_N}\right)^D \qquad (3-B)$$

This formula is adjusted for a strain range of the known point and the peak. It should be noted that there might be some unexpected deviations between the calculated and experimental flow stress for lower strains than the known point. Therefore, choosing the known point with the lowest strain is important in utilizing this method.

In order to show how this method works, I chose the data with the highest differences between the experiment and estimation, i.e. $\dot{\varepsilon}=1 s^{-1}$, and T=573 K (see "Appendix C"). Four different estimations are shown in Fig.1-B. The experimental value of peak stress is used in the estimations A and B, and the predicted one (see "Appendix C") for C and D. The other difference between these estimations is the value of $\sigma_P - \sigma_0$. The method explained in text, i.e. the extracted value from the x-intercept of the linear plot of $\ln(\sigma_P - \sigma)$ vs. $\ln\left(1-\dfrac{\varepsilon}{\varepsilon_P}\right)$ as $\ln(\sigma_P - \sigma_0)$, is used for estimations B and D. For A and C the method developed in this appendix is used. As it can be seen, the best fits belong to curves A and B, respectively. This indicates that the prediction of the peak stress intensely affects the results. Compared to other curves, curve D doesn't show a good accordance with the experiment.

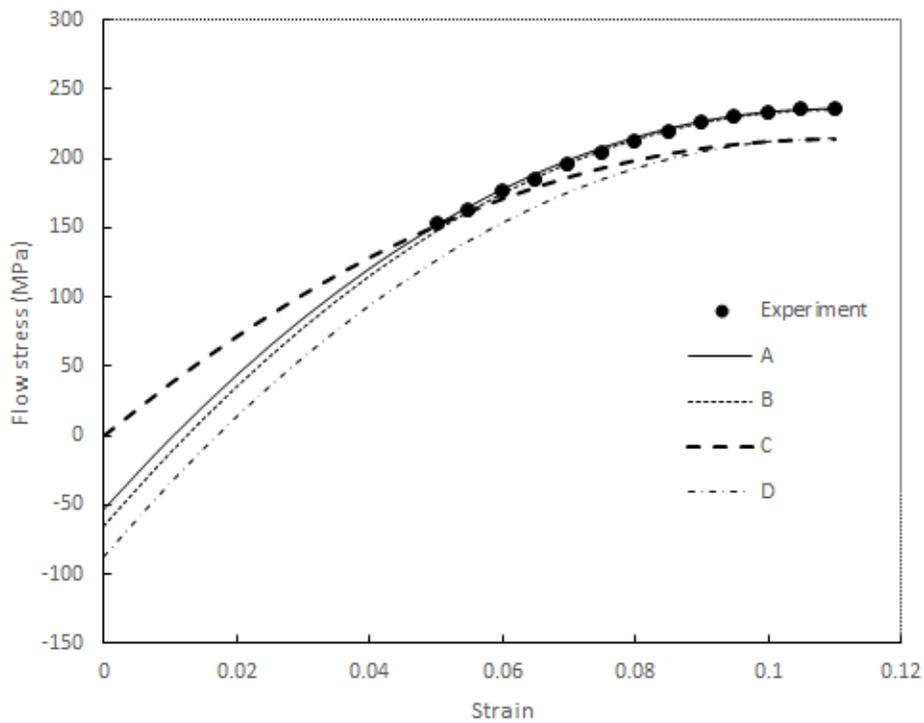

**Fig. 1-B.** Comparison between experimental and predicted flow stress (by Eq. 5) of a strain rate of 1 s$^{-1}$ and at a temperature of 573 K.

It should be noted that the value of the initial stress by the method developed in this appendix is just a hypothetical value (which, irrationally, might be a negative value) only for estimation of the flow stress at strains higher than the experimental data point with the lowest strain. Fig. 1-C also shows that using the method explained in this appendix results in a higher precision than the method explained in the text; i.e. curves A and B, respectively.

# Appendix C

The values of peak stress can be calculated from the Garofalo sine hyperbolic model [1]:

$$\dot{\varepsilon} = B(\sinh(\alpha\sigma))^n \exp\left(-\frac{Q_{def}}{RT}\right) \qquad (1-C)$$

where $R$ is the gas constant, $Q_{def}$ the apparent hot deformation activation energy, and $B$, $\alpha$ and $n$ are constants. In order to find the constant $\alpha$, first the following equations should be considered:

$$\dot{\varepsilon} = B'\sigma^{n'} \exp\left(-\frac{Q_{def}}{RT}\right) \qquad (2-C)$$

$$\dot{\varepsilon} = B'' \exp(\beta\sigma)\exp\left(-\frac{Q_{def}}{RT}\right) \qquad (3-C)$$

where $B'$, $B''$, $n'$ and $\beta$ are constants. For the low stress level ($\alpha\sigma<0.8$) and high stress level ($\alpha\sigma>1.2$), Eq.1-C would be estimated from Eqs.2-C and 3-C, respectively. The value of $\alpha$ is equivalent to $\beta/n'$. Taking the natural logarithm of both sides of Eqs.2-C and 3-C shows that the values of $n'$ and $\beta$ can be obtained from the slope of the line in $\ln\sigma$ vs $\ln\dot{\varepsilon}$ (Fig.1-C) and $\sigma$ vs $\ln\dot{\varepsilon}$ (Fig.2-C) plots, respectively, which are found to be 7.55784 and 0.08816. Consequently, $\alpha=0.0117$.

Taking the natural logarithm of both sides of Eq.1-C would result in Eq.4-C:

$$\ln \dot{\varepsilon} = \ln B + n \ln(\sinh(\alpha\sigma)) - \frac{Q_{def}}{RT} \qquad (4-C)$$

which concludes the slope of $\ln(\sinh(\alpha\sigma))$ vs $\ln \dot{\varepsilon}$ plot as the value of *n*. This plot is shown in Fig.3-C and determines *n*=5.213.

In order to find the value of activation energy, Eq.4-C should be rearranged as follows:

$$\ln(\sinh(\alpha\sigma)) = \frac{\ln B}{n} - \frac{\ln \dot{\varepsilon}}{n} - \frac{Q_{def}}{nRT} \qquad (5-C)$$

This equation shows that the slope of $\frac{1}{nRT}$ vs $\ln(\sinh(\alpha\sigma))$ plot would be the activation energy, which is calculated as 202.23kJ.mol$^{-1}$. Fig.4-C shows this plot (in scale of kJ.mol$^{-1}$, i.e. the unit of the horizontal axis is $\frac{1000}{nRT}$).

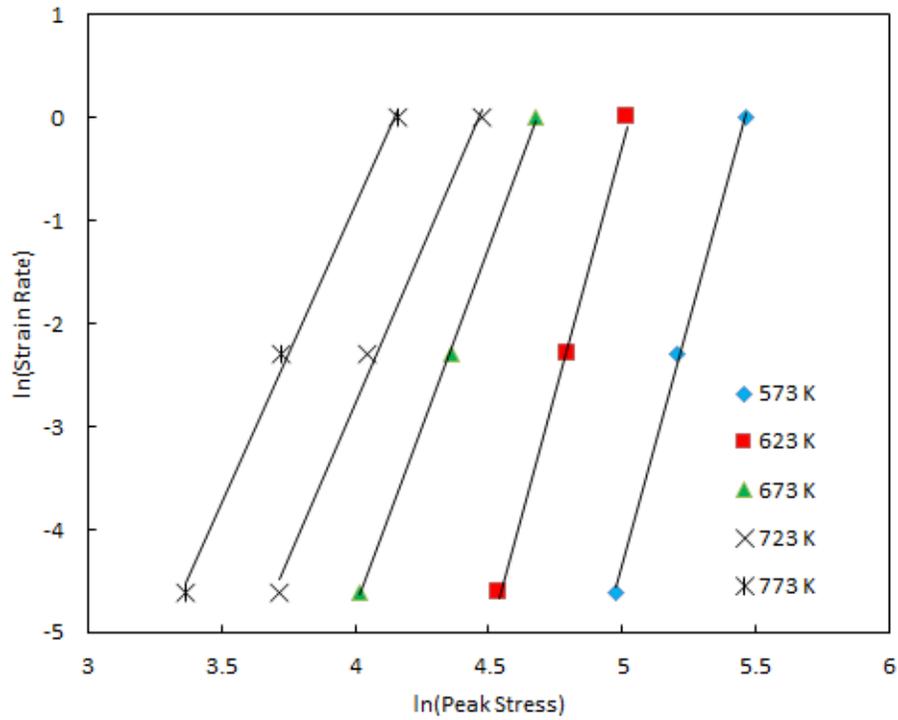

**Fig. 1-C.** Linear plots for determination of *n'*.

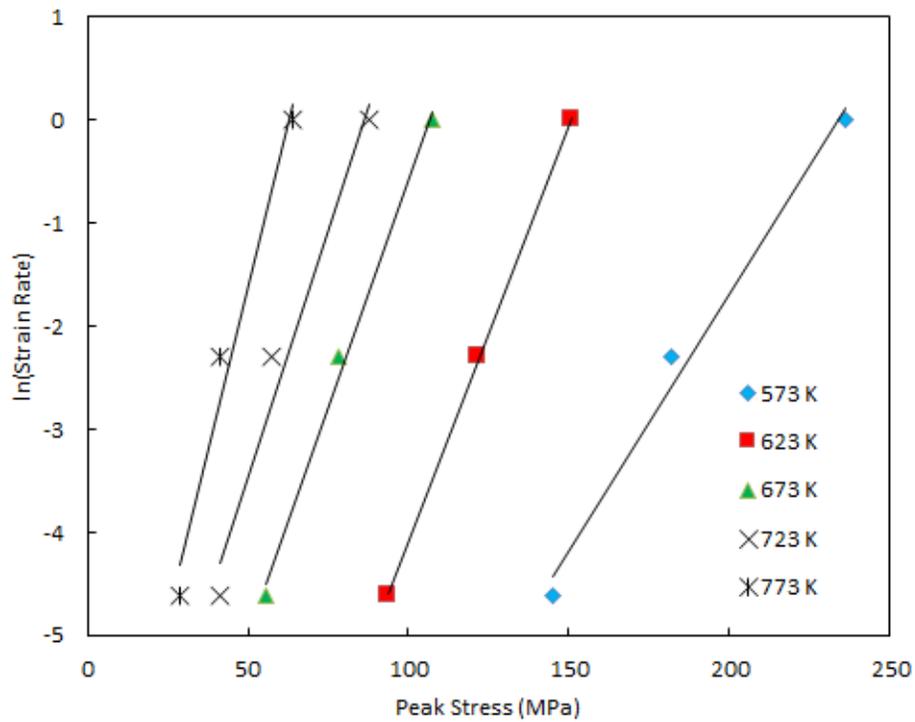

**Fig. 2-C.** Linear plots for determination of β.

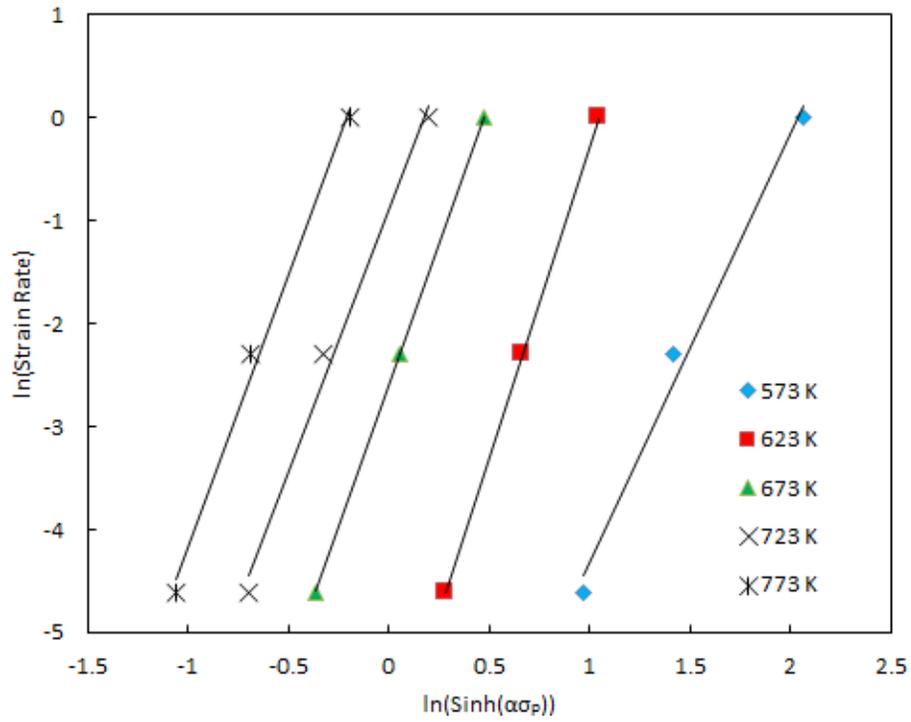

**Fig. 3-C.** Relationship between ln(strain rate) and ln(sinh(ασ$_P$)).

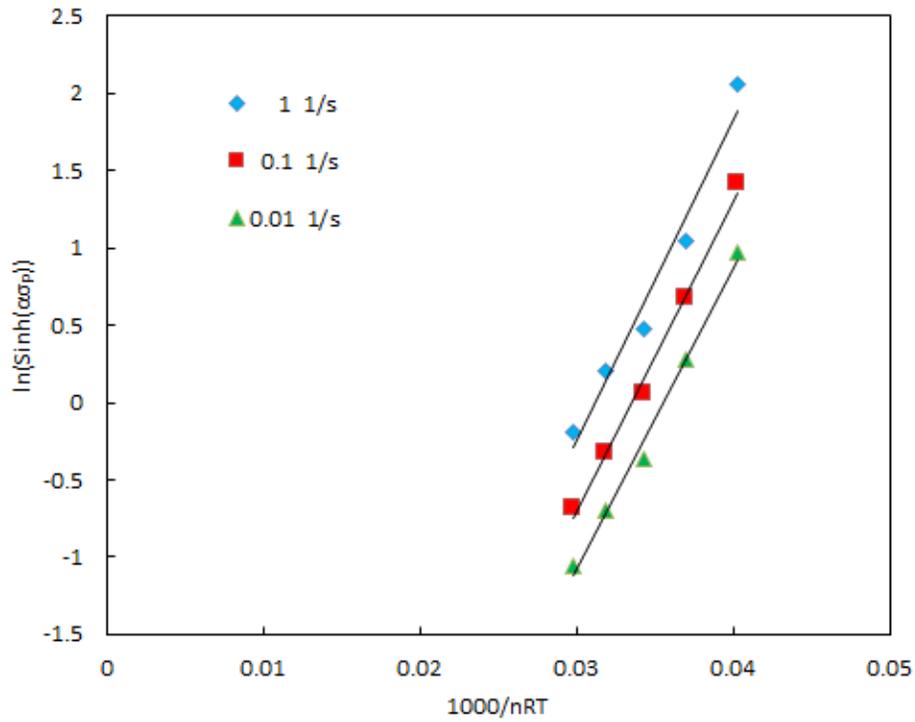

**Fig. 4-C.** Relationship between ln(strain rate) and 1000/nRT.

As the last step, constant B can be found from $\ln(\sinh(\alpha\sigma))$ vs $\ln\left(\dot{\varepsilon}\exp\left(\frac{Q_{def}}{RT}\right)\right)$ plot; the exponential of the y-intercept would be $B^{\ddagger}$, which is determined to be $\ln B = 32.02$. So, Eq.1-C for the peak stress would be:

$$\dot{\varepsilon} = \exp(32.02)(\sinh(0.0117\sigma_P))^{5.213}\exp\left(-\frac{202230}{8.314T}\right) \quad (6-C)$$

The comparison between the experimental and the calculated peak stress using Eq.6-C is illustrated in Fig. 5-C.

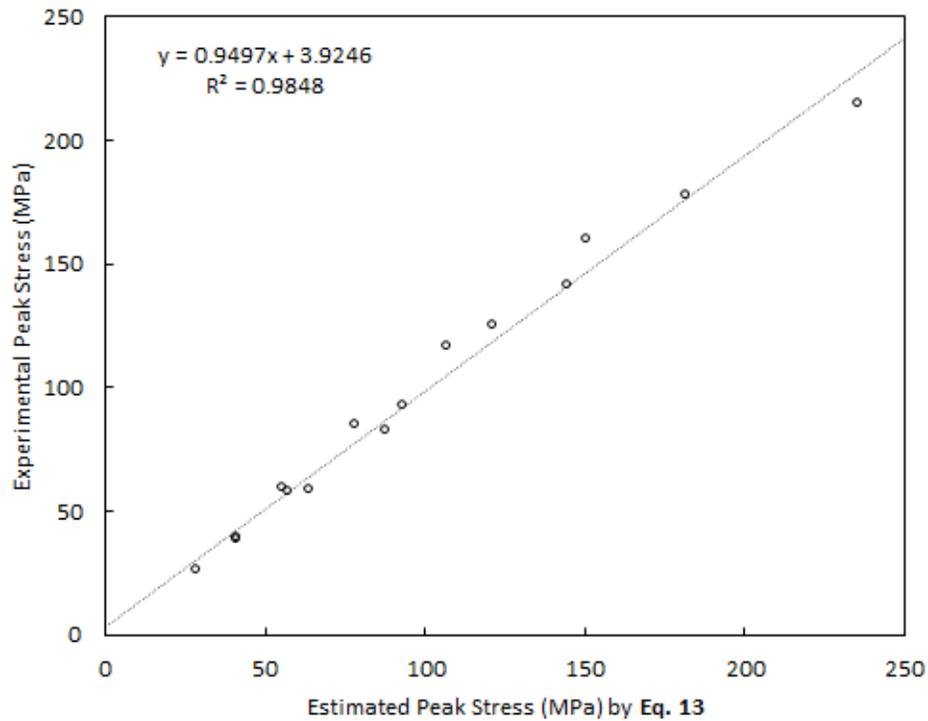

**Fig. 5-C.** Comparison between experimental and predicted peak stress.

The maximum, minimum, and average percent error between the experimental peak stress values and the estimated ones by Eq.6-C are 10.83, 0.56 and 5.59, respectively. Due to the errors of the predicted peak stress, using these values in calculation of the flow stress by Eqs.3 and 5 affect the results.

---

‡ The slope of the line MUST be equal to the calculated $n$.

Figs. 6-C and 7-C show the results of the predicted flow stress by Eqs. 3 and 5, respectively. The x-intercept of the linear plots of $\ln(\sigma_P - \sigma)$ vs. $\ln\left(1 - \dfrac{\varepsilon}{\varepsilon_P}\right)$ is used as the value of $\ln(\sigma_P - \sigma_0)$ in Eq. 3.

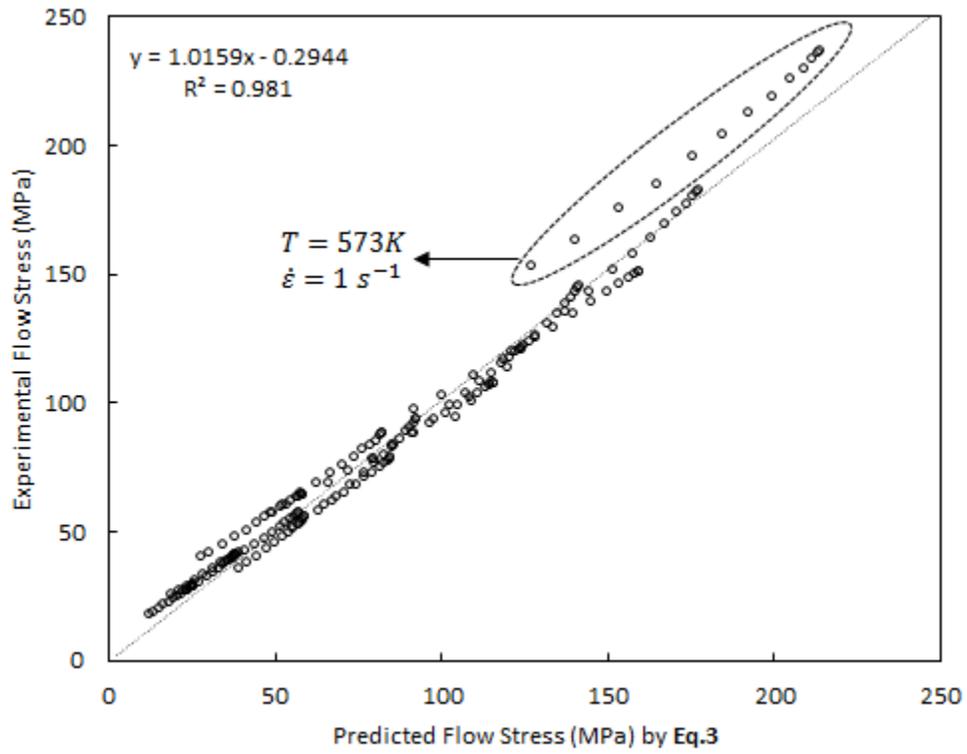

**Fig. 6-C.** Comparison between experimental and predicted flow stress by Eq. 3.

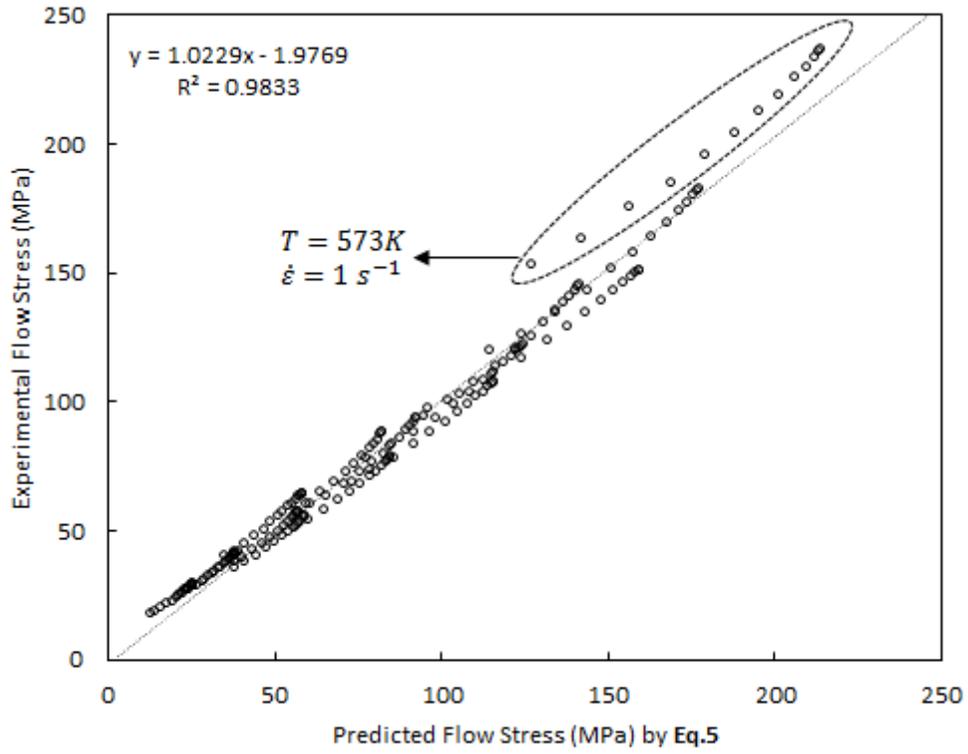

**Fig. 7-C.** Comparison between experimental and predicted flow stress by Eq. 5.

Using the value of the predicted peak stress in Eqs. 3 and 5, the average percent errors were 7.56 and 6.36, respectively.

# References

[1] F. Garofalo, An empirical relation defining the stress dependence of minimum creep rate in metals, Trans. Metall. Soc-AIME 227 (1963) 351–355.

[2] S. Solhjoo, Determination of critical strain for initiation of dynamic recrystallization, Mat. Des. 31 (2010) 1360–1364.

[3] S. Solhjoo, Determination of flow stress under hot deformation conditions, Mater. Sci. Eng. A 552 (2012) 566–568.

[4] H. Ahamed, V. Senthilkumar, Hot deformation behavior of mechanically alloyed Al6063/0.75Al2O3/0.75Y2O3 nano-composite—A study using constitutive modeling and processing map, Mater. Sci. Eng. A 539 (2012) 349–359.